\documentclass[conference]{IEEEtran}
\usepackage{graphicx}
\usepackage{amsfonts}
\usepackage{amssymb}
\usepackage[english]{babel}
\usepackage{amsmath,amsfonts,amssymb}
\usepackage{verbatim}
\usepackage{amsbsy}
\usepackage{algorithm}
\usepackage{algorithmic}
\usepackage[T1]{fontenc}		
\usepackage[utf8]{inputenc}		
\usepackage{hyperref} 
\usepackage{placeins} 
\usepackage{makeidx}
\usepackage[numbers,sort&compress]{natbib}
\begin{document}

\title{Tunable fiber source of entangled UV-C and infrared photons}

\author{\IEEEauthorblockN{$\textrm{Santiago Lopez-Huidobro}^{1,2,^*}$, $\textrm{Maria V. Chekhova}^{1,2}$}, $\textrm{Nicolas Y. Joly}^{2,1,3}$
\IEEEauthorblockA{$^1$ Max Planck Institute for the Science of Light, Staudtstraße 2, 91058 Erlangen, Germany}{$^2$ Department of Physics, Friedrich-Alexander-Universität, Staudtstraße 2, 91058 Erlangen, Germany}
\IEEEauthorblockA{$^3$ Interdisciplinary Centre for Nanostructured Films, Cauerstraße 3, Erlangen, Germany\\
$^*$ santiago.lopez-huidobro@mpl.mpg.de}
}


\maketitle

\thispagestyle{plain}
\pagestyle{plain}

\begin{abstract}
Pairs of entangled photons --- biphotons --- are indispensable in quantum applications. However, some important spectral ranges, like ultraviolet, have been inaccessible to them so far. Here, we use four-wave mixing in a xenon-filled single-ring photonic crystal fiber to generate biphotons with one of the photons in the ultraviolet and its entangled partner in the infrared spectral range. We tune the biphotons in frequency by varying the gas pressure inside the fiber and thus tailoring the fiber dispersion landscape. The ultraviolet photons are tunable from $271\,$nm to $235\,$nm and their entangled partners, from $764\,$nm to $1342\,$nm, respectively. The tunability up to $170\,$THz is achieved by adjusting the gas pressure by only $0.57\,$bar. At $1.32\,$bar, the photons of a pair are separated by more than $2$ octaves. The access to ultraviolet wavelengths opens the possibility for spectroscopy and sensing with undetected photons in this spectral range.
\end{abstract}

\IEEEpeerreviewmaketitle
\bigskip
Over the last few years, nonclassical light and in particular pairs of entangled photons (biphotons) have proved to be an indispensable tool for optical quantum technologies, especially quantum communication~\cite{gisin2007quantum}, metrology~\cite{giovannetti2011advances}, and sensing~\cite{lemos2022quantum}. In particular, considerable interest has recently emerged towards imaging and spectroscopy ``with undetected photons'', based on the induced coherence effect~\cite{zou1991induced,kalashnikov2016infrared,novikova2020study}. Due to the ``induced coherence'', one can perform imaging or spectroscopy of any material at the frequency of the probe photon by looking at the photon entangled to it, where the latter can be at a very different frequency. Thus, no detection is needed for the probe photons, offering a significant technological advantage as it can give access to challenging spectral regions, such as ultraviolet (UV) or extreme-UV (x-UV), which lack of good detection schemes. Many materials have absorption bands in the UV region and therefore generating biphotons with one of the photons in the UV, x-UV, or even at shorter wavelengths while its entangled partner is in a more accessible spectral range is particularly interesting to study. This however remains one of the open challenges of modern science and is still a subject of investigation~\cite{kutas2022quantum}. Biphoton generation in the UV-A was first reported in 2004 by Edamatsu  \textit{et~al.}~\cite{edamatsu2004generation}, who used biexciton resonant four-wave mixing (FWM) in a single semiconductor crystal. Nonetheless, the generated photons were only located $\sim2\,$THz around the frequency of the pump centered at $\sim390\,$nm. Moreover, the pump wavelength and the separation between the pair depended on the used material, strongly restricting the potential applications of such a source. A more common effect of generating biphotons, spontaneous parametric-down conversion (SPDC), requires a pump of a higher frequency than any of the daughter photons, which is difficult to implement. Some  SPDC experiments went as far into short wavelengths as to X-rays~\cite{danino1981parametric,borodin2017high}; however, no coincident photons have been reported so far.

In this work, we experimentally demonstrate a biphoton source using a gas-filled single-ring photonic crystal fiber (SR-PCF). The signal photon is tunable over the UV-C band, while the wavelength of the idler photon ranges from $764$ up to $1342\,$nm. The largest separation between photons of the generated pair is more than $2$ octaves. The tunability is achieved by adjusting the pressure of the filling gas. 

Our biphoton source relies on FWM in a xenon-filled SR-PCF, where the low nonlinearity of the gas is balanced by a large interaction length. FWM produces pairs of signal and idler photons provided that the phase-matching conditions hold:  
\begin{subequations}
\begin{align}
\omega_p - \omega_s &= \omega_i - \omega_p 
\label{FWMenergy},\\
2\beta_p  = \beta_s &+ \beta_i +2\gamma P_P ,  \label{FWMmomentum}
\end{align} \label{FWMeYm}
\label{E&M_cons}
\end{subequations}
which imply frequencies of the signal $\left(\omega_s\right)$ and of the idler $\left(\omega_i\right)$  photons to be symmetrically located around the frequency of the pump $\left(\omega_p\right)$. Here, $\beta_{(p,s,i)}$ are the propagation constants of the pump ($p$), signal ($s$), and idler ($i$), respectively. $\gamma$ is the nonlinear coefficient of the fiber and $P_P$ is the peak power of the pump. These conditions not only waive the necessity for exotic pump wavelengths but allow extreme frequency separation between the pair provided by the low dispersion over the considered bandwidth. In this regard, gas-filled anti-resonant hollow-core PCFs (HC-PCFs) are well adapted here. The group velocity dispersion of their fundamental mode is weakly anomalous at usual optical pump wavelengths. It can be balanced by the normal dispersion of the filling gas allowing exquisite control of the location of the zero-dispersion wavelength (ZDW). Combined with their ultrabroad transmission window, gas-filled HC-PCFs were already used to efficiently generate tunable ultraviolet bands~\cite{joly2011bright,jiang2015deep} by emission of dispersive waves, and also to engineer tunable Raman-free twin-beams of correlated photons in an HC-PCF filled with noble gases~\cite{finger2015raman,cordier2020raman,lopez2021fiber}. 

\begin{figure}[h!]
\centering
\includegraphics[width=1\linewidth]{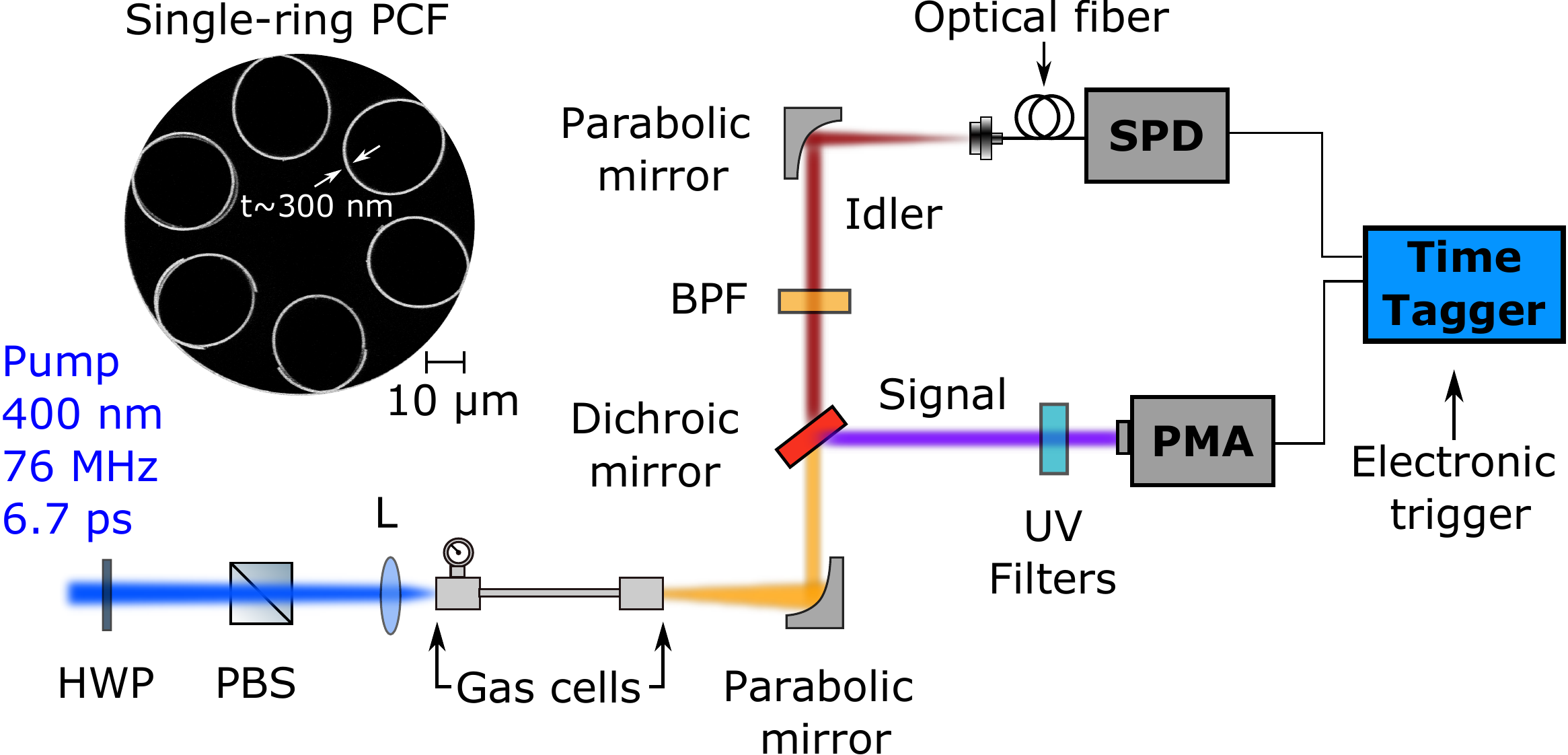}
\caption{Tunable biphoton source. The pump beam is coupled into the single-ring PCF with a lens L. A half-wave plate (HWP) and a polarizing beamsplitter (PBS) control the pump power. The output is collected with a parabolic mirror. The dichroic mirror separates the idler photon from the signal and the remaining pump photons. The pump radiation is suppressed with UV filters. Fluorescence is suppressed using an electronic trigger and a bandpass filter (BPF). Signal photons are detected with a photomultiplier detector assembly (PMA) and idler photons, with a single-photon detector (SPD) based on an avalanche photodiode. Two-photon coincidences are registered with a time-tagger. Inset: Scanning electron micrograph of the used single-ring PCF.}
\label{fig:Experimental_setup}
\end{figure}

Our biphoton source consists of a $65\,$cm long SR-PCF with a core diameter of $20.5\,\mu$m encircled by six silica tubes with an average diameter of $12\,\mu$m, and a thickness of $t\sim300\,$nm. A scanning electron micrograph of the used SR-PCF is shown in the inset of Fig. \ref{fig:Experimental_setup}. Based on anti-resonance guidance~\cite{zeisberger2017analytic, zeisberger2018understanding,vincetti2019simple}, the fiber was designed to guide the pump, signal, and idler photons with moderate losses at their fundamental spatial mode, with the resonances of the silica tubes being away from the aforementioned wavelengths. The spectral locations and the width of the fiber's resonances were calculated in a close analytical form, similar to~\cite{vincetti2019simple}.

\begin{figure}[h!]
\centering
\includegraphics[width=1\linewidth]{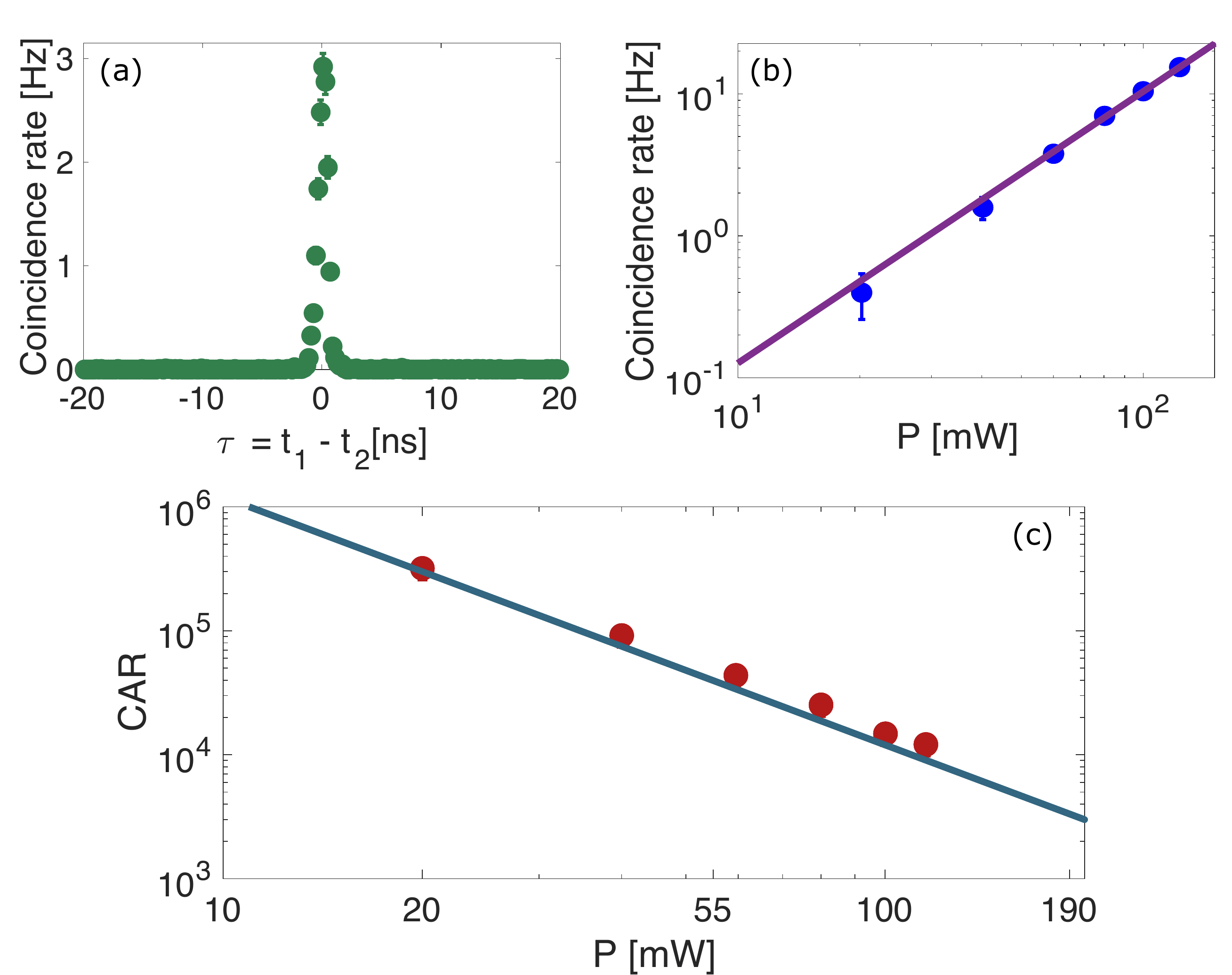}
\caption{(a) Histogram of the time delay between the detections of signal and idler photons. (b) Pump-power dependence of the coincidence rate. (c) Pump-power dependence of the coincidence-to-accidentals ratio (CAR). Measurements were performed at a constant pressure of $0.79\,$bar. Solid lines correspond to their respective fitting curves.}
\label{fig:fiber_charac}
\end{figure}

Both ends of the SR-PCF were mounted inside gas cells allowing the in-coupling of the light, while offering the possibility to fill and adjust the xenon pressure inside the fiber. The pump is a frequency-doubled Ti:Sapphire picosecond mode-locked oscillator operating at $400\,$nm with a repetition rate of $76\,$MHz. Signal and idler are respectively isolated with UV- and bandpass filters (BPFs) (see Methods \ref{Supplement1} for details) and measured by single-photon detectors connected to a time-tagger. The gate window for recording single events is set to $400\,$ps considering the timing jitter of both detectors. We trigger the opening of the recording window with the pulses from the pump laser. Figure \ref{fig:fiber_charac}(a) shows the distribution of the time interval $\tau = t_s-t_i$ between the arrivals of signal and idler photons, at a fixed xenon pressure of $0.79\,$bar. At this pressure, the wavelengths of the signal and idler are $\lambda_s=266\,$nm and $\lambda_i=800\,$nm. The pronounced peak centered at $\tau=0$ indicates the simultaneous arrival of both photons (two-photon coincidences) and the detection of biphotons. Due to the repetition rate of our source, we can expect accidental coincidences between photons generated by successive pulses with a time delay of $13.2\,$ns. Our results do not present any measurable evidence of such accidental coincidences at the corresponding delay (Fig.~\ref{fig:fiber_charac}(a)). Accidental coincidences originating from either fluorescence or pump photons could also contribute to the central peak. To demonstrate that such contributions are negligible in our setup, we measured the total coincidence rate as a function of the pump power (P) (Fig. \ref{fig:fiber_charac}(b)), where the total coincident rate was calculated by integrating the central peak. The results exhibit a quadratic behavior (fitted purple solid line), typical for biphoton generation via FWM~\cite{ivanova2006multiphoton, cordier2020raman}. 

To show the potential of our source to be implemented in quantum technologies, we estimated the coincidence-to-accidentals ratio (CAR): $\text{CAR} =g^{(2)}(0)-1$ where $g^{(2)} (\tau=0)$ is the normalized second-order correlation function,  
   \begin{equation}\label{normg2}
   g^{(2)}(\tau) = \frac{N_{si}(\tau)\text{R}_p}{N_s N_i},
\end{equation}
taken at zero delay. In Eq.\,(\ref{normg2}), $\text{R}_p$ is the repetition rate of the pulsed source, $N_{si}(\tau)$ is the rate of two-photon detection events, and $N_s$ (resp. $N_i$) is the single-count rate of the signal (resp. idler). To obtain the rate of FWM in each detection channel, we quantified the system's noise, which mainly originated from fluorescence and affected mostly the wavelengths longer than the pump (idler wavelength). We measured the amount of noise at different pump powers by isolating a spectral region without phase-matched idler photons. After subtracting the level of fluorescence on the idler channel, we estimated the CAR. Figure \ref{fig:fiber_charac}(c) shows CAR as a function of the pump power at a fixed pressure of $0.79\,$bar. The experimental points follow the $1/\text{P}^2$ dependence (fitted solid-teal line). The obtained high CAR values show potential for quantum technologies applications. The maximum value is $\text{CAR}_{\text{max}} \approx 3\times10^5$, which, to our knowledge, is among the best values reported to date in the literature for fiber-based biphoton sources.  

\begin{figure}[h!]
\centering
\includegraphics[width=1\linewidth]{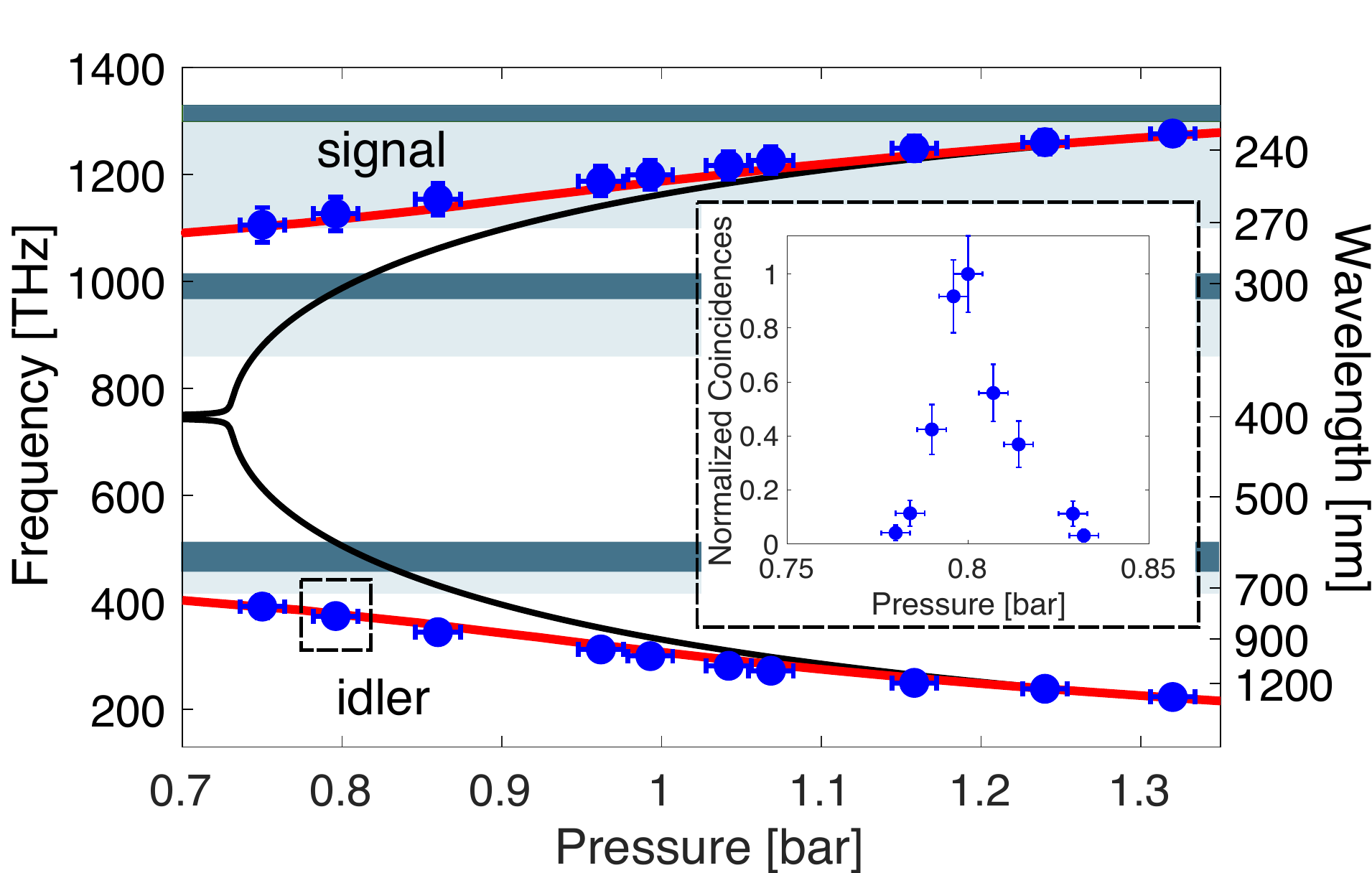}
\caption{Measured (points) and calculated (lines) wavelengths of the signal and idler photons as functions of the gas pressure inside the single-ring PCF. Red lines were calculated using Eqs.\,(\ref{E&M_cons}) with a thickness of $t=300\,$nm. Black lines were calculated using the model but ignoring the resonances~\cite{Marcatili:64}. The bandwidth of the used BPFs sets the vertical error bar. The teal bands indicate the location of the silica tubes’ resonances. Inset: pressure-dependent coincidence rate used to precisely locate the phase-matched wavelength, shown in the dashed square.}
\label{fig:tunabilitybiphotons}
\end{figure}

To tune the frequencies of the generated biphotons, we adjusted the pressure of the xenon filling the SR-PCF from $0.79$ up to $1.32\,$bar. At these pressures, the system remains in the normal dispersion regime, yielding well-separated signal and idler photons through FWM. Since our source operates at the single-photon level, we cannot directly use an optical spectral analyzer to measure the wavelengths of the generated biphotons. To measure the pressure-dependent spectrum, we first estimated the location of both signal and idler wavelengths at a given pressure by solving Eqs.\,(\ref{E&M_cons}), where the propagation constant for the fundamental linearly polarized $\text{LP}_{01}$ mode is given by the extended Marcatili-Schmertzel model~\cite{zeisberger2017analytic, zeisberger2018understanding} (red lines in Fig. \ref{fig:tunabilitybiphotons}; see Methods \ref{Supplement2} for details). We then selected the spectral region around the estimated location of the idler photon by inserting a BPF with a spectral bandwidth up to $40\,$nm. Subsequently, we measured the two-photon coincidence rate versus the xenon pressure, fine-tuned around the calculated value (inset of Fig. \ref{fig:tunabilitybiphotons}). The conjugated signal wavelength was calculated from the energy conservation. The blue dots in Fig.~\ref{fig:tunabilitybiphotons} show the pressure-dependent spectrum of the biphoton source. We repeated this method to measure the wavelengths of the photons for different pressures of xenon, with the resolution given by the spectral bandwidth of the used BPFs. 

We observe a giant spectral separation between the photons, with the signal photons being in the UV-C and the idler photons from the near-IR to the IR spectral range. The spectral separation between the photons increases along with the pressure up to more than two octaves achieved at $1.32\,$bar, where the signal and idler photons are at $235$ and $1342\,$nm, respectively. The spectral tunability is $\sim30\,$THz$/0.1\,$bar, amounting to $\sim170\,$THz for a pressure range of $0.57\,$bar.  

We stress that the perfect agreement between the experimental and theoretical data is due to using the extended Marcatili-Schmeltzer model~\cite{Finger:14, Marcatili:64}, accounting for the resonances, whose location is shown by horizontal dark-teal lines in Fig. \ref{fig:tunabilitybiphotons}. The presence of the cladding resonances strongly modifies the phase-matching curve, especially in the vicinity of the ZDW at $\lambda=422\,$nm. For comparison, the black line in Fig. \ref{fig:tunabilitybiphotons} indicates the phase-matching curve without considering those resonances. 

Meanwhile, the rate of biphoton generation is $440$ times less than estimated from the theory~\cite{palmett2009propiedades}.  We believe this is caused by the inhomogeneity of the silica tubes along the fiber length. As their thickness changes, so do the locations of the cladding resonances where photons are no longer guided. We confirmed that the thickness of the silica tubes might range from $300$ up to $340\,$nm by measuring the transmittance of the fiber (see Methods \ref{Supplement3}). Figure \ref{fig:tunabilitybiphotons} shows the spectral range of the resonances considering the inhomogeneity of the silica tubes marked with horizontal light-teal bands. Since only the guided photons can eventually be detected, we presume that only the last $3\,$cm of the fiber contributed to the measured biphotons.

\begin{figure}[h!]
\centering
\includegraphics[width=1\linewidth]{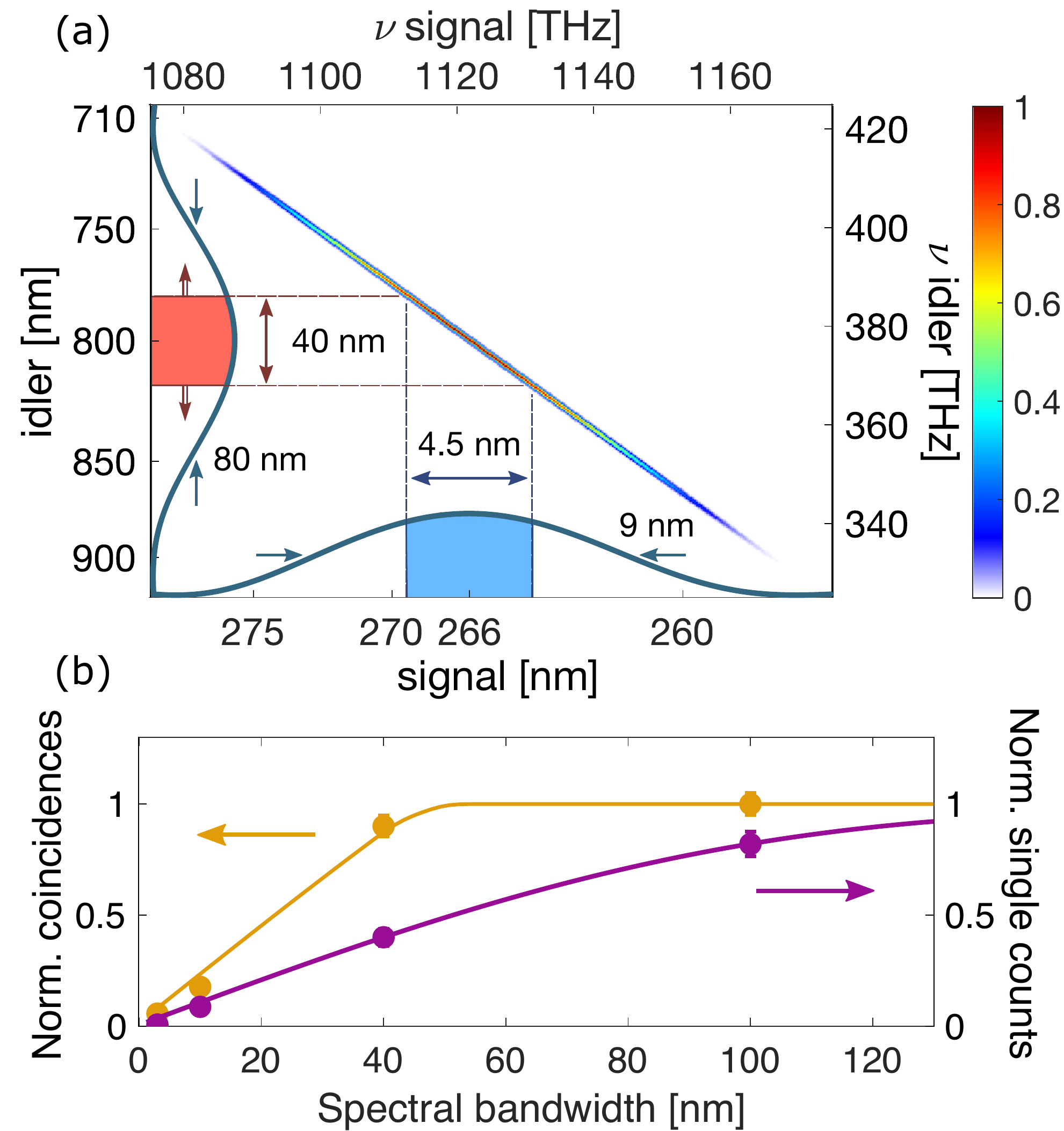}
\caption{(a) Calculated JSI at a pressure of $0.79\,$bar. Teal curves show the spectra of signal and idler photons. Vertical and horizontal bands indicate the filtered spectral bands for signal and idler photons, in the coincidence measurement. (b) Normalized rates of coincidences (yellow) and single counts (purple) measured (points) and calculated (lines) as functions of the selected idler’s spectral bandwidth.}
\label{fig:JSI}
\end{figure}

Finally, we study the joint spectrum of biphotons emitted by our source. Figure \ref{fig:JSI}(a) shows the joint spectral intensity (JSI) for a pressure of $0.79\,$bar calculated similarly to~\cite{garay2008tailored}. The anti-diagonally stretched shape indicates spectral anti-correlations between the signal and idler photons. The projections of the JSI on the signal and idler axes (teal curves) reveal the spectra of the corresponding photons with a full-width at half maximum of $80\,$nm for the idler and $9\,$nm for the signal photons. To verify the frequency anticorrelations between the generated biphotons, we compared the ``conditional'' spectrum of the idler photons, obtained by registering coincidences under narrowband ($4.5\,$nm, depicted in Fig. \ref{fig:JSI}(a)) filtering of the signal photons, and their ``unconditional'' spectrum, obtained by measuring single counts. Based on the calculation, the ``conditional'' distribution should be much narrower than the unconditional one, and this is confirmed by the experiment. Indeed, as we increase the spectral bandwidth in the idler channel (Fig. \ref{fig:JSI}(b), see Methods \ref{Supplement1} for details), the coincidence rate (yellow) ``saturates'' already at a spectral bandwidth of $\sim40\,$nm, which roughly corresponds to collecting all idler photons matching the bandwidth selected in the signal channel. Meanwhile, the rate of single counts (purple) increases up to the maximal bandwidth of $100\,$nm, which indicates the broad ``unconditional'' spectrum of idler photons. The experimental points agree with the theoretical curves,  obtained by integrating the respective spectra of the signal and idler photons. These results indirectly demonstrate the frequency anti-correlations between the signal and idler photons since they show how the correlated spectrum was restricted by selecting a fraction of the signal photon spectrum only. 

In conclusion, we have experimentally demonstrated the first  biphoton source with one of the photons in the UV-C and its entangled partner in the near-IR spectral range. The frequencies of signal and idler photons are tunable without changing the pump frequency or the fiber, but by simply varying the gas pressure inside the SR-PCF~\cite{finger2017characterization}. We tuned these frequencies over $\sim170\,$THz by adjusting the pressure from $0.79\,$ to $1.32\,$bar. At the highest pressure, the spectral separation between the photons exceeded $2$ octaves. Finally, at a fixed pressure of $0.79\,$bar, we estimated the JSI along with the spectral bandwidths of the generated biphotons and experimentally verified the spectral entanglement of the pair. 

This source will be useful for quantum spectroscopy or other types of sensing with undetected photons, and for developing sensor devices in the UV spectral range at the single-photon level. In particular, due to the use of UV signal photons, improved resolution is expected in the case of ``imaging with undetected photons''~\cite{fuenzalida2022resolution}. The access to the UV-C spectral range, which this novel biphoton source provides, paves the way toward a technological revolution with ``undetected photons'' techniques.

\bigskip

\section{Methods}

\subsection{Experimental details of the tunable biphoton source}\label{Supplement1}

\begin{figure}[h!]
\centering
\includegraphics[width=0.9\linewidth]{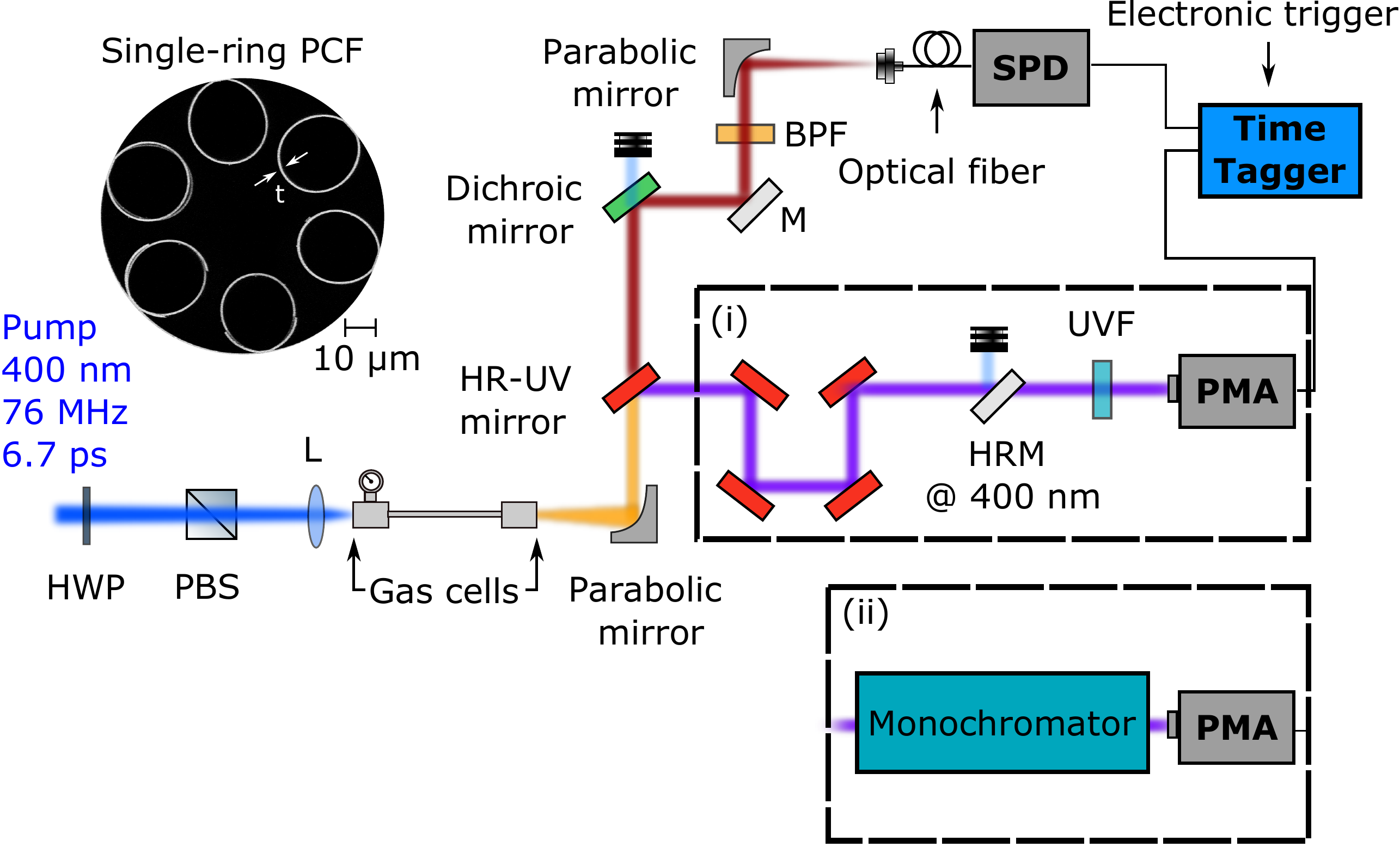}
\caption{Tunable biphoton source. The pump beam is launched into the single-ring PCF with a lens L. A half-wave plate (HWP) and a polarizing beamsplitter (PBS) control the pump power. The output is collected with a parabolic mirror. High-reflective UV mirrors (HR-UV) separate the signal and idler photons. The pump radiation is suppressed with four HR-UV mirrors, two 400 nm-high-reflective mirrors (HRM), and a UV filter (UVF). The signal photons are detected with a photomultiplier detector assembly (PMA). The idler photons are detected with a single-photon detector (SPD), based on an avalanche photodiode,  after suppressing the residual pump with a dichroic mirror and a band-pass filter (BPF). Two-photon coincidences are recorded with a time-tagger. Inset: Scanning electron micrograph of the used single-ring PCF.}
\label{fig:Experimental_setup_supplement}
\end{figure}

The initial pump centered at $400\,$nm has a pulse duration of $6.7\,$ps and a repetition rate of $R_p=76\,$MHz. We launched the pump into the SR-PCF with a lens L with a focal length $f=~100\,$mm. The output beam is collected with an $f=56.5\,$mm parabolic mirror. Subsequently, the signal and idler photons are spectrally separated with a long-pass dichroic mirror (cut-off wavelength $275\,$nm). This mirror is highly reflective for the UV spectral range and has $99\,\%$ reflectivity at $255\,$nm (HR-UV mirror). To filter out the pump, the signal photons are reflected successively with four HR-UV mirrors (see inset (i) of Fig. \ref{fig:Experimental_setup_supplement}).

For the detection part, the signal photons are registered with a photomultiplier detector assembly (PMA), sensitive from $185\,$nm to $820\,$nm with a maximum quantum efficiency of $20\,\%$ at $250\,$nm. To further suppress the pump, two $400\,$nm-highly-reflective mirrors (HRM @ $400\,$nm) and a band-pass colored glass filter centered at $310\,$nm (UVF) were located in front of the PMA (see inset (i) of Fig. \ref{fig:Experimental_setup_supplement}). The transmitted idler photons are reflected with a short-pass dichroic mirror (cut-on wavelength $625\,$nm) to separate the remaining pump spectrally. Finally, with the help of silver-coated (M) and parabolic mirrors, the idler photons are fed into a single-mode optical fiber (SMF-28 commercial fiber) and eventually detected with a single-photon detector (SPD). To cover the near-IR and IR spectral ranges, silicon- and InGaAS/InP-based SPDs were used accordingly. The corresponding photon-detection events from each detector translate into digital output pulses, which are sent to a time-tagger to register the single-count rates of the signal and idler photons ($N_s$, $N_i$), and the rate of coincidence detection events ($N_{si}$) between the two channels. 

To verify the spectral correlations of the generated biphotons, we performed a “conditional” two-photon coincidences. In this experiment, we replaced the optics from inset (i) see Fig. \ref{fig:Experimental_setup_supplement}, with a self-made monochromator built for the UV-C spectral range (see inset (ii) of Fig. \ref{fig:Experimental_setup_supplement}). This measurement consisted in recording two-photon coincidences by selecting, with the help of the monochromator, $4.5\,$nm bandwidth of the central part of the signal photon’s spectrum. On the IR channel, we varied the spectral bandwidth of the idler photon from $3$ to $100\,$nm using different bandpass filters. At each respective IR spectral bandwidth, a coincidence measurement was done. Simultaneously, for each different bandpass filter, the single-count photons from the IR channel were measured. The system's noise, mainly originating from fluorescence, was subtracted as described in the main manuscript. All the experiments were performed at a fixed pressure of $0.79\,$bar and with an average pump power of $140\,$mW.

\subsection{Dispersion of the xenon-filled single-ring PCF}\label{Supplement2}

In the extended version of the Marcatili-Schmertzel model, the propagation constant for the fundamental linearly polarized $\text{LP}_{01}$ mode is given by~\cite{zeisberger2017analytic, zeisberger2018understanding}

\begin{equation}
n_{\text{eff}} = n_{\text{gas}} - \frac{j^2_{0,1}}{2 k_0^2n_{\text{gas}}R^2} - \frac{j^2_{0,1}}{k_0^3n^2_{\text{gas}}R^3}.\frac{\text{cot}\left[\Psi(t) \right]}{\sqrt{\epsilon -1}}.\frac{\epsilon + 1}{2}.
\label{effRefIndex}
\end{equation}
Here, the function $\Psi(t)=k_0 t \sqrt{n_{si}^2-n_{\text{gas}}^2}$, $R$ is the fiber’s radius, and $\epsilon=n_{si}^2/n_{\text{gas}}^2$ is the ratio of the refractive index of the silica tubes ($n_{si}$) to those of the filling gas ($n_{\text{gas}}$). The silica tubes surrounding the core have a thickness $t$. $j_{(0,1)}=2.405$ is the first zero of the zero-order Bessel function of the first kind. This extended version of the Marcatili-Schmertzel model for a capillary includes the pressure-dependent contribution of the filling gas $k_0 n_{\text{gas}}(P_r)$ and the influence of the waveguide ($2^{\text{nd}}$ term). Additionally, it includes the presence of resonances ($3^{\text{rd}}$ term), where the light can couple out from the core into the cladding, and the locations of these non-guiding regions depend on $t$.

\subsection{Characterization of the single-ring PCF}\label{Supplement3}

\begin{figure}[h!]
\centering
\includegraphics[width=0.80\linewidth]{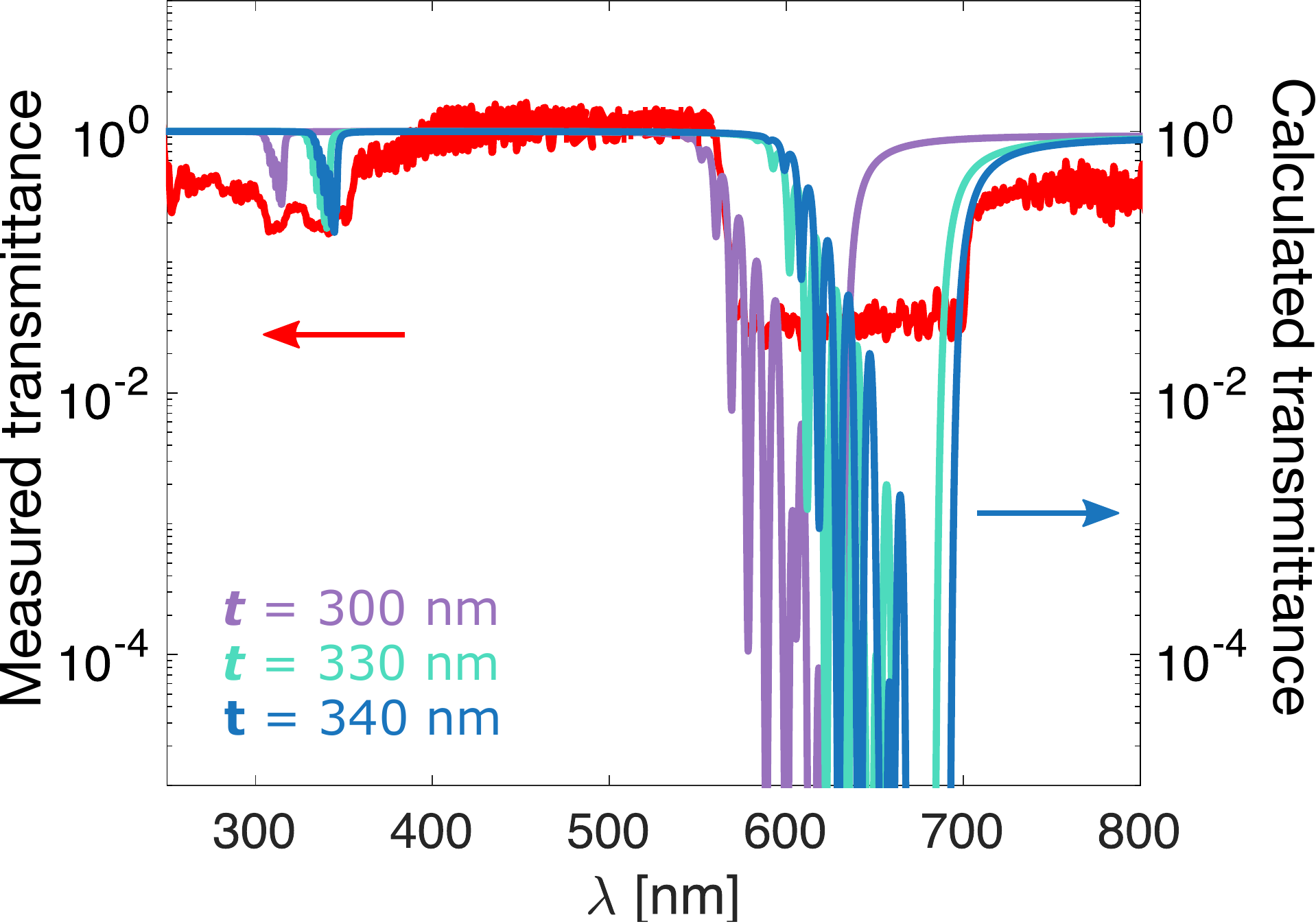}
\caption{Measured (red) and calculated transmittance (purple, green, and blue) of the used single-ring PCF.}
\label{fig:fiber_characterization}
\end{figure}

To estimate the thickness of the silica tubes inside the SR-PCF, we measured the transmittance of a $30\,$cm-long fiber pumped with a self-made supercontinuum source, extending from $280$ up to $850\,$nm. At first, we measured the transmission spectrum with the whole fiber length. Consecutively, we cut the fiber leaving only $3\,$cm and measured the transmission spectrum again. With the latter, we took into account the light-coupling loss and precisely measured the pump’s spectrum at the entrance of the fiber. The transmittance (red curve) was deduced from both spectra and it is shown in Fig. \ref{fig:fiber_characterization}. The resonances of the silica tubes appear in the form of strong degradation of the transmittance.

In Fig. \ref{fig:fiber_characterization} we added the calculated transmittance using~\cite{vincetti2019simple}. The different colors correspond to the different thicknesses of the silica tubes. Since the experimental measurement corresponds to the estimated transmittance using several thicknesses, we can conclude that the thickness of the silica tubes mainly varies from $300$ to $340\,$nm along the total length of the fiber. 

\bigskip 
\textbf{Funding} We acknowledge the financial support by Deutsche Forschungsgemeinschaft (DFG) (Grants No. CH-1591/9-1 and No. JO-1090/6-1). 
\bigskip

\textbf{Disclosures} The authors declare no conflicts of interest.

\bigskip

\textbf{Data availability} Data underlying the results presented in this paper are not publicly available at this time but may be obtained from the authors upon reasonable request.

\bigskip 

\bibliography{sample}

\end{document}